# van der Waals heterostructures based on atomically-thin superconductors

*Carla Boix-Constant, Samuel Mañas-Valero\*, Rosa Córdoba, Eugenio Coronado\**

Universidad de Valencia (ICMol), Catedrático José Beltrán n° 2, Paterna 46980, Spain.

E-mail: samuel.manas@uv.es, eugenio.coronado@uv.es

Keywords: Van der Waals heterostructures, 2D materials, transition metal dichalcogenides, charge density waves, superconductivity, electronic transport

**Abstract**

Van der Waals heterostructures (vdWHs) allow the assembly of high-crystalline two-dimensional (2D) materials in order to explore dimensionality effects in strongly correlated systems and the emergence of potential new physical scenarios. In this work, it is illustrated the feasibility to integrate 2D materials in-between 2D superconductors. Particularly, it is presented the fabrication and electrical characterization of vertical vdWHs based on air-unstable atomically-thin transition metal dichalcogenides formed by $NbSe_2/TaS_2/NbSe_2$ stacks, with $TaS_2$ being the insulator $1T-TaS_2$ or the metal $2H-TaS_2$. Phase transitions as $1T-TaS_2$ charge density wave and $NbSe_2$ superconductivity are detected. An enhancement of the vdWH resistance due to Andreev reflections is observed below the superconducting transition temperature of the $NbSe_2$ flakes. Moreover, in the $NbSe_2$ superconducting state, the field and temperature dependence of the normalized conductance is analyzed within the Dynes´ model and the overall behavior is consistent with the Bardeen-Cooper-Schrieffer theory. This vdWH approach can be extended to other 2D materials, such as 2D magnets or topological insulators, with the aim of exploring the new emergent properties that may arise from such combinations.

# 1. Introduction

Van der Waals heterostructures (vdWHs) offer the opportunity to assemble high quality two-dimensional (2D) materials with different physical properties in order to tune their performance or explore the new emergent physical behaviors that can arise.[1] Beyond the first graphene-based vdWHs and the remarkable observation of superconductivity in twisted bilayer graphene,[2] transition metal dichalcogenides (TMDCs) –layered materials formed by the sequential stacking of X-M-X planes, where X is a chalcogen and M is a transition metal **(Figure 1)**– have played a fundamental role in the development of the emergent field of the van der Waals 2D materials.

Among TMDCs, the most studied vdWHs are based on the semiconducting group VI (M = Mo, W) compounds due to its electronic and optoelectronic properties.[3] Much less studied are the vdWHs composed by group V (M = V, Nb, Ta) TMDCs, probably due to its fast oxidation in air. In fact, only few examples involving mainly thin-layers of $NbSe_2$ have been reported in the literature.[4] In bulk, group V TMDCs have attracted large interest due to its strongly correlated phenomena such as several charge density waves (CDWs) configurations, superconductivity and even quantum spin liquid (QSL) phases. In particular, bulk 2H-$NbSe_2$ is a superconductor with a superconducting critical temperature ($T_c$) of 7.2 K and CDW at 33 K, bulk 2H-$TaS_2$ is a superconductor with a $T_c$ of 0.8 K and CDW at 75 K and bulk 1T-$TaS_2$ is ascribed as a Mott insulator and a QSL candidate with a CDW formation at 550 K, a near-commensurate CDW (N-CDW) to an incommensurate CDW (I-CDW) transition at 350 K and a commensurate CDW (C-CDW) to N-CDW transition at 200 K.[5] In the case of 1T-$TaS_2$, the relationship between the out-of-plane stacking of the CDW and its magnetic properties is currently under debate.[6] Below 200 K, the CDW forms the so-called Star-of-David arrangement in bulk 1T-$TaS_2$, where every 13 Ta atoms are coupled (12 of them pair and shift towards the central one, making a frustrated triangular lattice of $S = ½$ electrons). If out-of-plane correlations are absent, 1T-$TaS_2$ should be a metal since there is an odd number of electrons per unit cell. Thus, the Mott mechanism has been invoked to explain its insulating behavior and it is the basis for the existence of a quantum spin liquid. However, if the Star-of-David are coupled in the out-of-plane direction forming dimers, 1T-TaS$_2$ must be considered as a band insulator since there would be an even number of electrons per unit cell and the quantum spin liquid picture would not be valid anymore.[7] In fact, recent experimental works have shown that the CDW forms out-of-plane dimers at low temperatures –thus, being a band insulator and not a QSL– but, at higher temperatures, the system transits to a Mott insulating state – compatible with a QSL–.[8] Moreover, current theoretical models have manifested the importance of the out-of-plane stacking of the CDW: whereas the formation of dimers yields to an insulating ground state, the stacking through the vertical or diagonal directions –A and L configurations, following the notation of ref. [9]– favors a metallic one.[10] Nonetheless, other scenarios such as the formation of emergent domain wall networks,[11] the out-of-plane dimerization[12] or the Mott picture,[8] among others, [6] are also under consideration.

Thus, the use of different TaS$_2$ polytypes allow the study of different physical backgrounds ranging from a semiconductor/insulator for 1T-TaS$_2$ to a metal/superconductor in the case of 2H-TaS$_2$. In addition, when dealing with thin-layers of these materials drastic changes in their properties due to dimensionality effects can be observed.[13] For instance, there is a $T_c$ suppression in 2H-NbSe$_2$[14] but a $T_c$ enhancement in 2H-TaS$_2$,[15] being currently under debate the CDW role in the 2D limit. Therefore, vdWHs based on thin-layers of TMDCs offer a unique platform for studying strongly correlated materials in reduced dimensionality systems.

Moreover, vdWHs based on superconducting materials separated by a thin barrier (for example, a structure like NbSe$_2$/TaS$_2$/NbSe$_2$, **Figure 1.e** and **Figure 1.f**) offer new perspectives in the field of vertical junctions, both from the experimental and theoretical points of view. Superconducting heterostructures based on atomically-thin vertical junctions have been important for studying tunnel junctions and Josephson junctions since their discovery by Giaever back in the 60's.[16] Compared with these pioneering works,[17] current vdWHs exhibit new experimental differences. For example, new materials and fabrication techniques are involved –from junctions prepared by metal evaporation to mechanically stacked flakes forming vdWHs– and the theoretical modelling becomes more complex –from type-I superconductors as aluminum or lead to type-II ones like $NbSe_2$ with the competition of CDWs–.

In this work, the electronic transport properties of vdWHs based on thin-layers of 2H-$TaS_2$ and 1T-$TaS_2$ sandwiched between superconducting thin-layers of $NbSe_2$ (**Figure 1**) are inspected. Although in the present vdWHs the geometrical factors cannot be fully controlled, since mechanical exfoliation yields to flakes with random shapes, the present heterostructure approach benefits from being fully integrated under inert atmosphere conditions, thus paving the way to work with air-unstable 2D materials like $NbSe_2$ or $TaS_2$ layers that, otherwise, would not be possible to explore due to its fast oxidation under ambient conditions.[18]

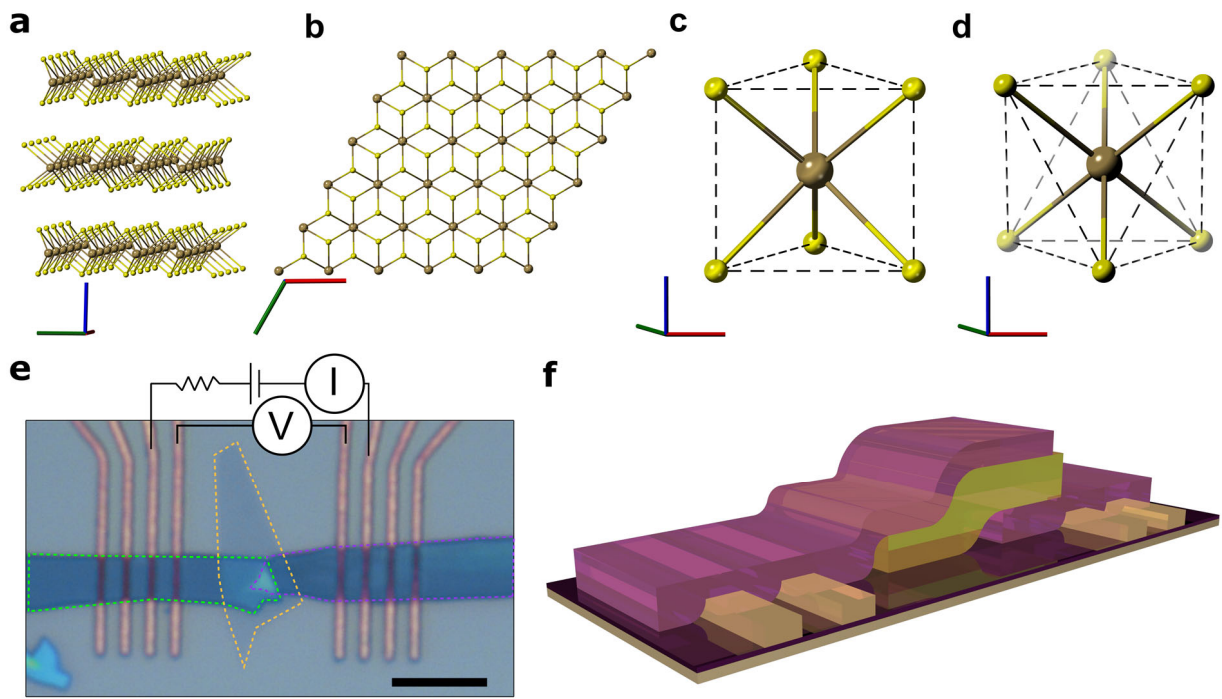

**Figure 1.-** van der Waals heterostructures (vdWHs) based on transition metal dichalcogenides (TMDCs). a) Out-of-plane layered structure of a 2H-TMDC. b) In-plane hexagonal structure (view along the c axis) of a 2H-TMDC. c) Detail of the 2H polytype. d) Detail of the 1T polytype. e) Optical image of a vdWH based on a $TaS_2$ thin-layer (yellow dashed lines) sandwiched by a top and bottom $NbSe_2$ thin-layer (green and purple dash-lines) with the electrical transport configuration sketched. Scale bar: 5 µm. f) Artistic vision (not to scale) of the vdWH. $NbSe_2$ layers are represented in purple, $TaS_2$ layer in yellow and the contact pads in gold. In a-d, the chalcogen atoms are shown as yellow balls and the metal ones as brown balls. The red, green and blue colors of the axis correspond to the a, b and c axis, respectively.

## 2. Results and discussion

vdWHs are fabricated by the deterministic stacking of mechanically exfoliated flakes, as previously reported (see Methods for further details).[19] All the process is done inside an argon glove box since thin-layers of $NbSe_2$ and $TaS_2$ degrade in air. We note that special care has to be taken with grounding and manipulating the devices since, by our experience, $NbSe_2$-based heterostructures suffer much more electrostatic discharges than graphene-based ones. Optical images, geometrical parameters and transport measurements of all the measured vdWHs are shown in the Supplementary Information. A total of 14 vdWHs were fabricated. 6 of them showed similar comprehensible behaviors (3 based on $1T-TaS_2$ and 3 based on $2H-TaS_2$) and are shown in the Supplementary Information. The rest suffered electrostatic discharges or were too resistive, thus making not possible to characterize its electrical properties up to 2 K (see, as an example, Figure S.54). This higher resistance can be attributed to a too thick barrier as a consequence of a small junction area, especially for the $1T-TaS_2$ case, which gives rise to non-linear IV curves even above the superconducting transition of the $NbSe_2$ flakes (see, as an example, device B in the Supplementary Information), or a bad contact at the interfaces.

In **Figure 2**, we plot the temperature dependence of the vdWHs resistance together with the temperatures where different transitions in the bulk have been reported for semiconducting $1T-TaS_2$ (N-CDW to CWD transition at 350 K and C-CDW to N-CDW at 200 K), metallic $2H-TaS_2$ (CDW at 75 K) and metallic $NbSe_2$ (CDW at 33 K and superconductivity at 7.2 K ).[5]

Let us first discuss the properties of the $NbSe_2/1T-TaS_2/NbSe_2$ heterostructure (**Figure 2.a**). At 350 K, this vdWH exhibits the typical hysteresis from the I-CDW to N-CDW of $1T-TaS_2$. Above this temperature, it exhibits a metallic dependence whereas below it behaves as a semiconductor. In the studied vdWHs geometry, the in-plane and out-plane components of $TaS_2$ can contribute to the overall resistance of the vdWH, together with the barrier formed at the $NbSe_2/TaS_2$ interface. However, the small resistance change observed in the CDW hysteresis at 350 K —when compared with previous reports in the literature for horizontal devices of 1T-TaS$_2$ thin-layers[20] and in bulk 1T-TaS$_2$[6]— suggest that the in-plane properties of 1T-TaS$_2$ are not playing the main role in the vdWHs behavior (see **Supplementary Section 2**). As well, the N-CDW to C-CDW hysteresis of bulk 1T-TaS$_2$ at 200 K is absent, as already reported in 1T-TaS$_2$ thin-layers and attributed to dimensionality effects.[18.a] While cooling down in the semiconducting range, a progressive non-linear transition is observed without any sharp or abrupt discontinuity until a more pronounced slope is reached below the superconducting critical temperature of the NbSe$_2$ flakes, as it is discussed in more detail below. This is in contrast with the out-of-plane behavior observed in bulk 1T-TaS$_2$, where clear linear slopes are reported.[6] Based on previous works (see introduction), our results would be in agreement with a gradual formation of out-of-plane dimers of the CDW structure, with a progressive change in the ratio between QSL layers and dimerized ones while cooling down.[21] The overall transport properties of the vertical vdWH can be modelled following an Arrhenius model, where the conductance (G) is of the form $G = G_0 e^{-\frac{E_a}{k_B T}}$ being G$_0$ a prefactor, E$_a$ the activation energy, k$_B$ the Boltzmann constant and T the temperature. As detailed in **Supplementary Section 3**, two regimes are studied: one in the high temperature range (N-CDW region, *i.e.*, 200 K – 350 K, where no dimerization would be expected) and one at low temperatures (T < 10 K, where dimerization is most likely to occur). The following activation energies are obtained: 5 ± 3 meV for the high temperature region and 0.4 ± 0.5 meV for the low temperature one. These values are comparable with the ones previously reported: 10 meV in the case of 1T-TaS$_2$ bulky samples[22] and 11 ± 7 meV (0.05 ± 0.03 meV) for vdWHs based on few-layers graphene (FLG) in the high (low) temperature range.[23] The higher activation energies at lower temperatures observed now can be attributed to a larger Fermi level mismatch between NbSe$_2$/1T-TaS$_2$ if compared with FLG/1T-TaS$_2$. Due to the multiple CDW transition that are presented in 1T-TaS$_2$, variable range hoping (VRH) can be as well considered as a possible transport mechanism, as already discussed in other 2D materials or vertical molecular junctions.[24] In this case, the conductance takes the form $G = G_0 \exp\left(-\frac{T_0}{T}\right)^x$, where $G_0$ is the residual conductance (that can exhibit a temperature dependence or not, see **Supplementary Section 7** for further details), $T_0$ is the characteristic hopping temperature and x is the hopping exponent that determines the scaling behavior. [24] As detailed in **Supplementary Section 7**, we have considered the 2D-VRH, 3D-VRH and nearest-neighbor-hopping (NNH) transport mechanism with both temperature-dependent and temperature-independent $G_0$ term. We note that, although none of the models give rise to a single transport mechanism all over the whole temperature range, the best fittings are obtained for the 2D-VRH case. Considering the 2D-VRH case with a temperature-dependent $G_0$ term, two crossovers are found at *ca.* 40 K and at 350 K (I-CDW). Regarding the 2D-VRH case with a temperature-independent $G_0$ term, multiple crossovers are found that roughly coincide with temperatures where some electronic transitions have been previously described for bulk 1T-TaS$_2$. As a consequence of the complex electronic structure of 1T-TaS$_2$, it is not possible to unambiguously unveil the most determinant transport mechanism at the moment.

As far as the properties of the NbSe$_2$ / 2H-TaS$_2$ / NbSe$_2$ heterostructure are concerned (**Figure 2.b**), a metallic behavior is observed upon cooling down until the top NbSe$_2$ flake transits to the superconducting state at 6.45 K ($T_c$ defined as a 50 % reduction of the resistance; see also **Figure 3**). At this temperature the resistance of the vdWHs drops significantly as a consequence of this superconducting transition until the bottom NbSe$_2$ flake also transits to the superconducting state (at 6.1 K). Below this temperature a sharp enhancement in the resistance is observed as a consequence of Andreev reflections (this is discussed below with more detail). In our thinnest devices, a second resistance drop at temperatures close to 2 K is also observed (see **Supplementary Section 1.6** and **1.7**). However, due to our base temperature of 2 K, it is not possible to conclude if this second transition is due to the superconducting transition of the 2H-TaS$_2$ flake or to the possible formation of a Josephson junction. For unraveling this point, future experiments at milliKelvin temperature will be carried out. Similar trends were observed for the different vdWHs measured (see **Supplementary Section 1**). We note that the resistivity values at room temperature of the NbSe$_2$ / 2H-TaS$_2$ / NbSe$_2$ vdWHs are comparable to the NbSe$_2$ / 2H-TaS$_2$ / NbSe$_2$ ones (see **Supplementary Section 5**), despite the different electronic nature of 1T-TaS$_2$ and 2H-TaS$_2$. This fact is compatible with measuring the out-of-plane component of 2H-TaS$_2$ due to the larger anisotropy between the out-of-plane and in-plane resistivity of 2H-TaS$_2$, that is enhanced in the thin-layer limit.[15]

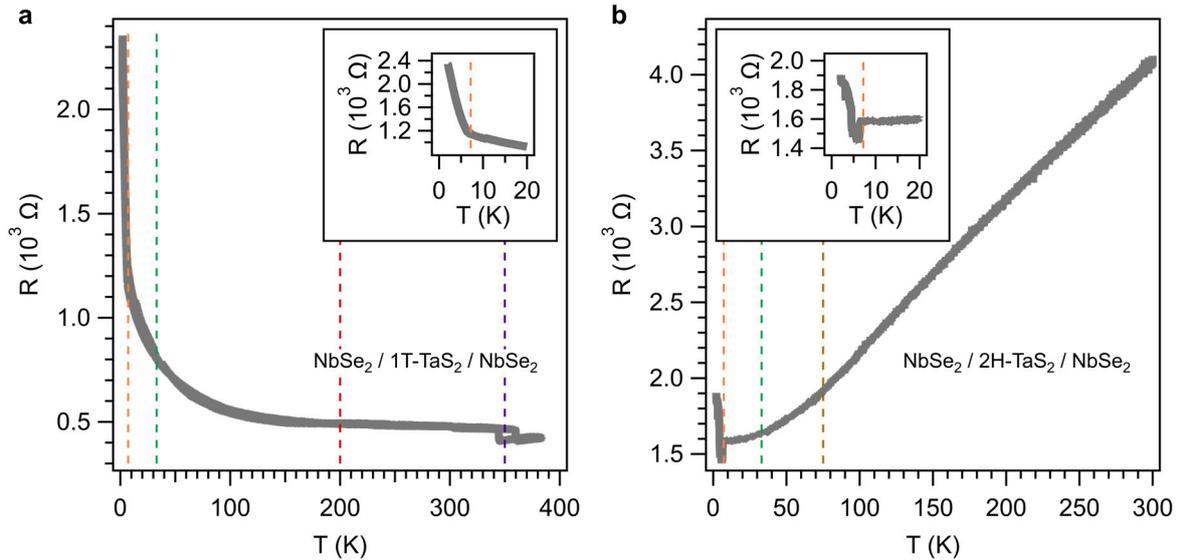

**Figure 2.-** Resistance ($I_{DC}$ = 1 µA) vs. temperature of the vdWHs: a) NbSe$_2$/1T-TaS$_2$/NbSe$_2$ vdWH (device A in the Supplementary Information) and b) NbSe$_2$/2H-TaS$_2$/NbSe$_2$ vdWH (device D in the Supplementary Information). Previous transitions temperatures reported in the literature for bulk NbSe$_2$ and TaS$_2$ are marked with orange (7.2 K, superconducting transition of NbSe$_2$), green (33 K, CDW of NbSe$_2$), red (200 K, C-CDW to N-CDW transition of 1T-TaS$_2$), purple (350 K, N-CDW to I-CDW transition of 1T-TaS$_2$) and brown (75 K, CDW of 2H-TaS$_2$) vertical dashed lines. The low-temperature regime is shown in the insets.

The low temperature properties of the two types of vdWHs have been investigated with more detail by performing DC IV curves with a small AC voltage, thus accessing as well to the differential resistance, dV/dI (see Methods). A representative example of the observed behavior is provided by the NbSe$_2$/2H-TaS$_2$/NbSe$_2$ vdWH (device D in the Supplementary Section). We show in **Figure 3** the temperature dependence of the resistance in the 2K – 8 K range at zero external magnetic field. As mentioned above, a resistance drop is observed when the top NbSe$_2$ contact enters into the superconducting state, as seen by performing DC IV curves and fitting them in the ohmic range (**Figure 3.a and 3.d**), until a significant increase in the resistance is observed when both top and bottom NbSe$_2$ contacts are superconducting. In the same line, the differential resistance, dV/dI (**Figure 3.b**), switches from a straight line when both NbSe$_2$ flakes are in the normal state (metallic) to a zero-bias peak in the superconducting state, with a decrease in the intensity when warming up. For NbSe$_2$/NbSe$_2$ heterostructures, a supercurrent below Tc has been observed.[25.a] In our case, this supercurrent is absent although the superconducting effects of NbSe$_2$ are still noticeable. Thus, in **Figure 3.c** the normalized conductance (see **Supplementary Section 4**) exhibits a deep around zero bias below T$_c$ followed by two-symmetric maxima (in blue) developing while cooling down. Similar trends have been observed in graphene junctions with conventional superconducting electrodes.[25.b] The conductance curves can be fitted by the phenomenological Dynes´ model [26] in order to estimate an energy gap Δ (see details in **Supplementary Section 4**). The temperature dependence of Δ (**Figure 3.e**) can be adjusted to the Bardeen–Cooper–Schrieffer (BCS) theory, yielding Δ(0) = (0.948 ± 0.005) meV, in good agreement with the values reported for thin-layers of NbSe$_2$,[27] and T$_c$ = (5.93 ± 0.02) K. The superconducting gap to T$_c$ ratio, 2Δ/k$_B$T$_c$, where k$_B$ is the Boltzmann constant, is 3.7, very close to the BCS value, 3.53.[28]

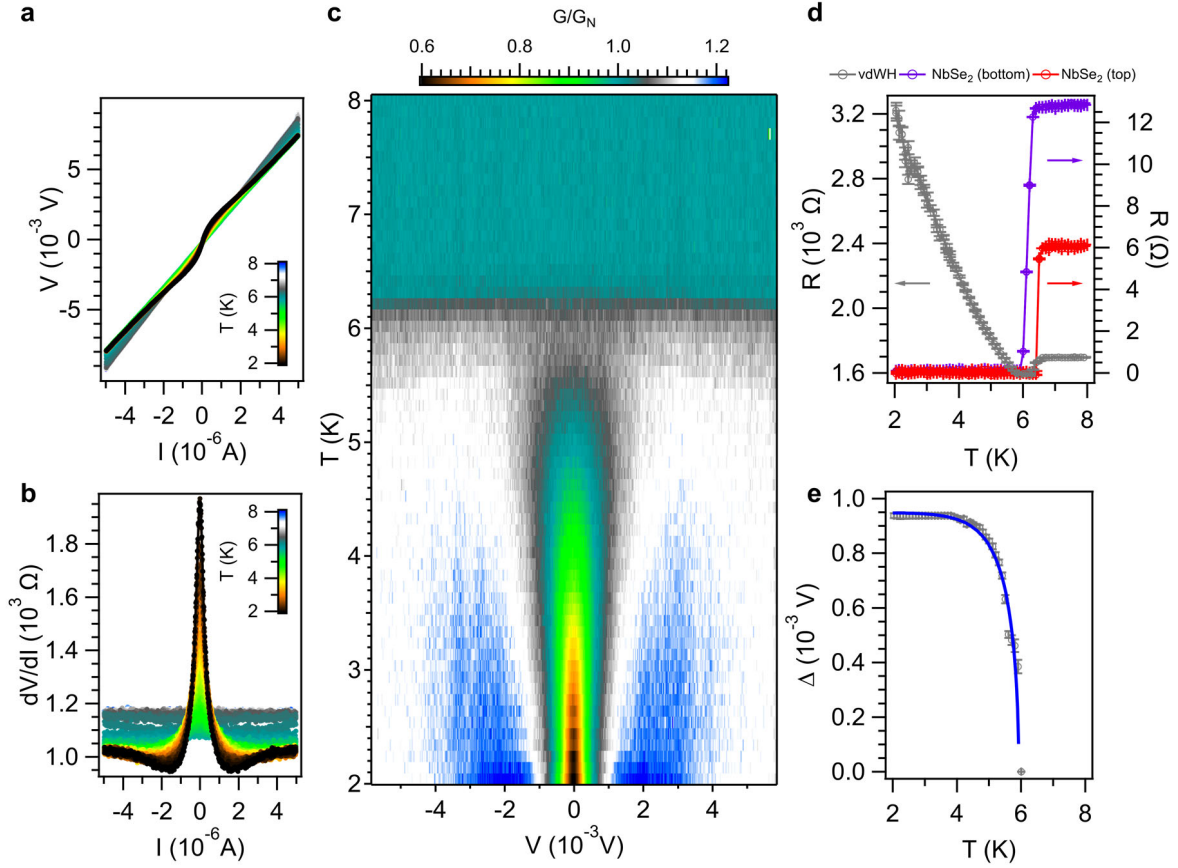

**Figure 3.-** Electronic transport properties of a NbSe$_2$/2H-TaS$_2$/NbSe$_2$ vdWH at low temperatures (device D in the Supplementary Information): a) DC IV curves from 2 K to 8 K, b) AC differential resistance from 2 K to 8 K, c) normalized conductance as a function of DC voltage bias and temperature, d) thermal dependence of the resistance obtained by fitting in the ohmic regime the DC IV curves of the vdWH and the top and bottom NbSe$_2$ flakes, e) thermal dependence of the energy gap obtain from fitting the curves shown in c to the Dynes´ function together with the BCS fitting.

The transport properties at 2 K of the NbSe$_2$/2H-TaS$_2$/NbSe$_2$ vdWH (device D in the Supplementary Section) under an external applied magnetic field (in the -8 T – 8 T range) perpendicular to the silicon substrate are shown in **Figure 4.** The resistance of the vdWH decreases as the field is increased while the NbSe$_2$ resistance is zero; then, the vdWH resistance increases with a slope change at the upper critical field. This trend can be seen from DC IV curves (**Figure 4.a**) and the corresponding fitting in the ohmic regime (**Figure 4.d**). In accordance, there is a suppression of the zero bias peak observed in the differential resistance (**Figure 4.b**) and of the two symmetric maxima around 0 voltage bias in the normalized conductance (in blue in **Figure 4.c**) while applying a magnetic field. The energy gap obtained by fitting the normalized conductance (**Figure 4.c**) to the Dynes´ function (see **Supplementary Section 4**) exhibit a parabolic dependence with the magnetic field (**Figure 4.e**), following $\Delta(B) \propto \sqrt{1 - \frac{B}{B_{c2}}}$.[29] From the fitting, it is obtained $B_{c2} = (2.655 \pm 0.004)$ T that, within the Ginzburg-Landau framework, yields to a coherence length $\xi_{GL} = \sqrt{\frac{\Phi_0}{2\pi B_{c2}}} \sim 11$ nm, where $\Phi_0$ is the quantum of magnetic flux.[28] Similar trends have been observed in different vdWHs (see **Supplementary Information**). The main difference between 1T-TaS$_2$ and 2H-TaS$_2$ based heterostructures is the absence in the former of the resistance drop near the critical temperature and critical field of the NbSe$_2$ flakes. This may be attributed to the more resistive behavior of the 1T-TaS$_2$ flakes.

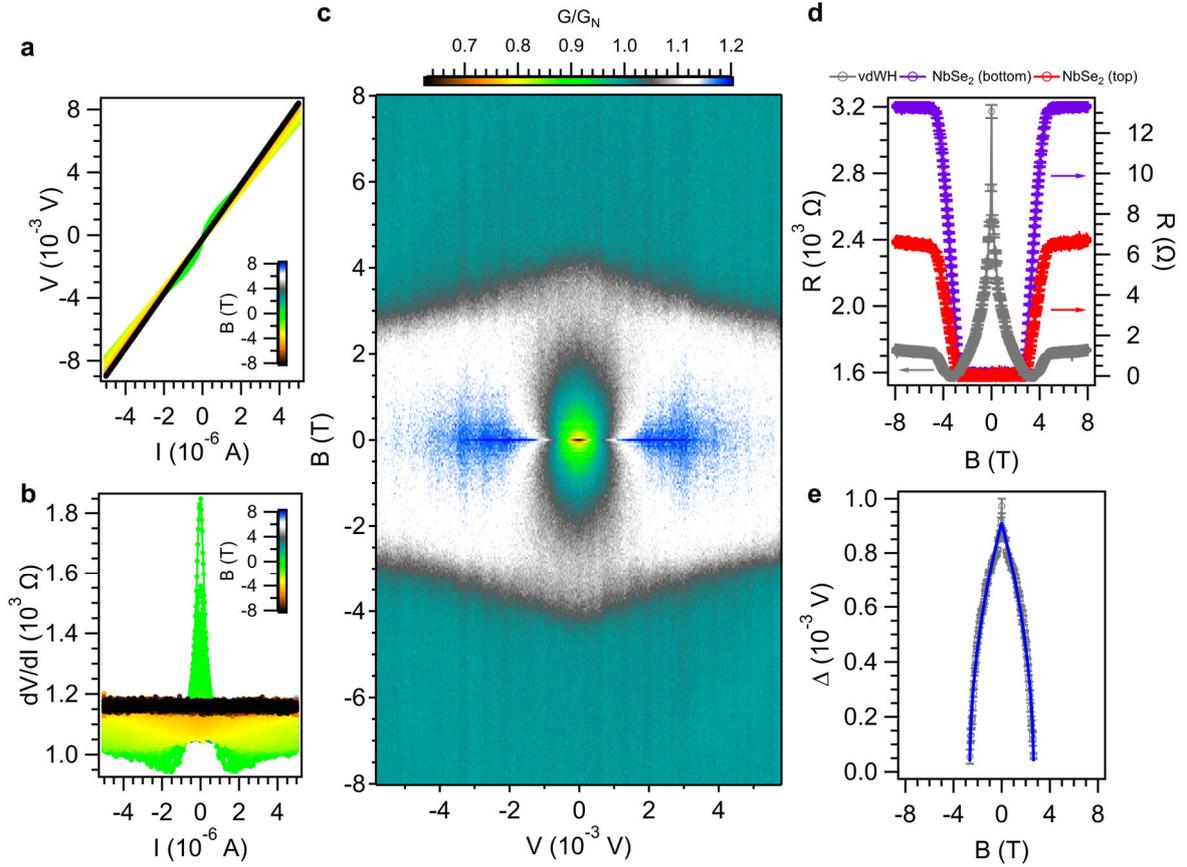

**Figure 4.-** Electronic transport properties of a NbSe$_2$/2H-TaS$_2$/NbSe$_2$ vdWH under an applied external magnetic fields at 2 K (device D in the Supplementary Information): a) DC IV curves from – 8 T to 8 T, b) AC differential resistance from -8 T to 8 T, c) normalized conductance as a function of DC voltage bias and external magnetic field, d) field dependence of the resistance obtained by fitting in the ohmic regime the DC IV curves of the vdWH and the top and bottom NbSe$_2$ flakes, e) field dependence of the energy gap obtain from fitting the curves shown in c to the Dynes´ function with the corresponding fitting.

Summarizing, in the vdWHs based on the superconducting thin layers reported above we have observed an enhancement of the resistance when the NbSe$_2$ layers become superconducting. Although this result may seem contra-intuitive, it can be understood as a consequence of Andreev processes. Let us discuss this aspect in the following. Above T$_c$, NbSe$_2$ is a normal metal (N) and the overall transport can be modelled by considering the band structure of the different elements involved, plus the barriers formed at the interfaces due to Fermi level mismatching.[30] This mismatching can be large – as determined, for example, by the formation of a Schocttky barrier when using insulators (I) or large band gap semiconductors, being the transport dominated by electrons or holes (**Figure 5.a**)– or small –for instance, when a normal metal is used as barrier (**Figure 5.b**)–.[28] These interfaces within a normal metal can be seen as a source of electron scattering where energy, spin and charge must be conserved, but not the linear momentum.[31] This is not the case below T$_c$, where energy, spin and momentum must be conserved, but not charge. This process is known as Andreev reflection (AR).[31] In the superconducting ground state, the charge is coupled forming Cooper pairs, yielding to the formation of a superconducting energy gap. Thus, an AR involves the transfer of a charge 2e at the interface (being e the elementary electron charge); the formation of a Cooper pair in the superconductor from an incident electron (hole) implies that a hole (electron) is retro-reflected in the normal metal with opposite spin but equal momentum (**Figure 5.c**).[30] In our case the increase in the resistance observed below the critical temperature and the critical field can be understood as a consequence of the formation of an energy barrier at the interface due to the AR process. Interestingly, when two superconductors are forming the heterostucture, the so-called Andreev bound states (ABS) can be produced. In this case resonant electron-hole states are formed in the central conductor (**Figure 5.d**). The two maxima observed in our case in the normalized conductance may arise from these ABS (**Figure 3.c** and **Figure 4.c**). More experiments such as microwave measurements or the use back gate voltages will be performed in the future to check this point.[33]

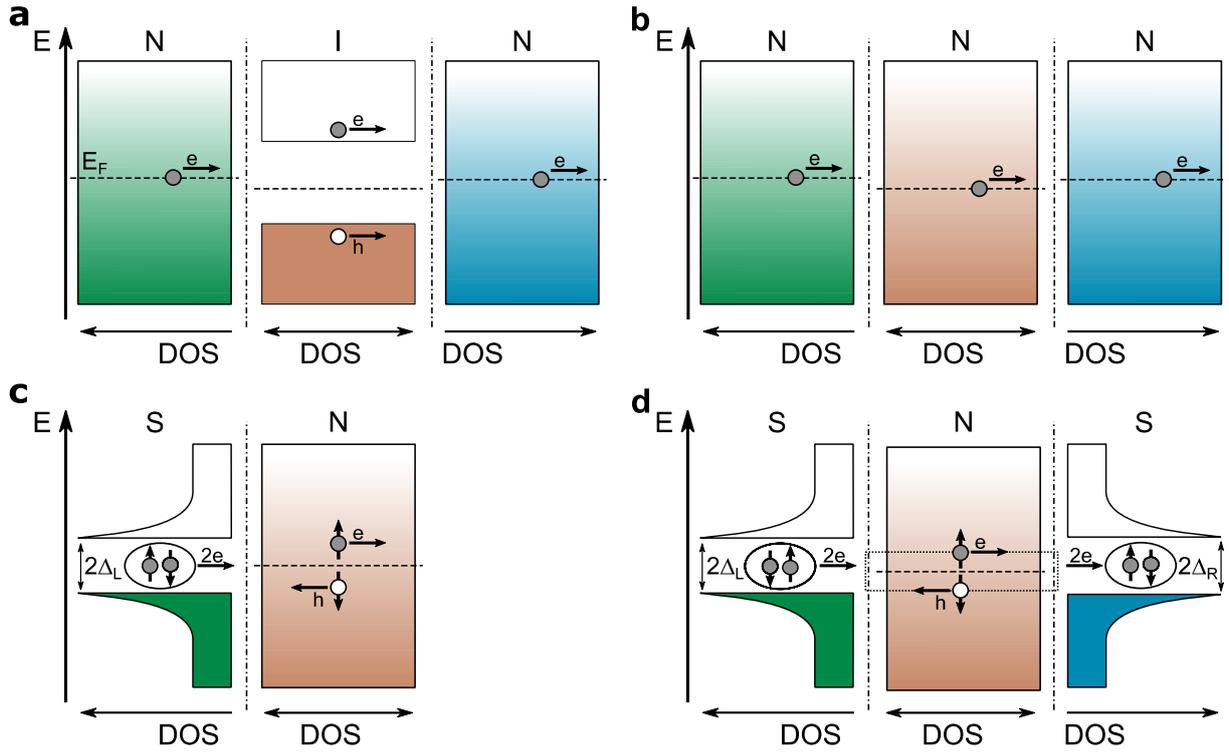

**Figure 5.-** Schematic band representation (energy versus density of states, DOS) of a) NIN (large interface barrier), b) NNN (low interface barrier or ohmic barrier), c) SN (Andreev reflection) and d) SNS (Andreev bound state, represented with points) vdWHs. Normal metals are denoted as N, insulators as I, superconductors as S and the interfaces as vertical dash-dotted lines. Fermi level is represented as a horizontal dashed line. Electrons (holes) are symbolized as grey (white) circles, where horizontal (vertical) arrows show its momentum (spin). The formation of Cooper pairs is symbolized as two electrons with opposite spin inside a circle, where $\Delta$ correspond to the superconducting energy gap and L (R) to the left (right) superconductor.

## 3. Conclusion

In this work, we have reported on the fabrication and electrical transport characterization of vertical van der Waals heterostructures based on atomically thin-layers of different $TaS_2$ polytypes, featuring semiconducting or metallic properties, in between superconducting $NbSe_2$ layers. In the area of the 2D materials, this constitutes the first attempt to form vdWHs integrating different air unstable transition metal dichalcogenides. This approach has allowed us to detect CDW in semiconducting 1T-$TaS_2$ and superconductivity in $NbSe_2$. In addition, we have observed at low temperatures —in the superconducting state of the NbSe$_2$ flakes— an increase in the resistance of the different vdWHs, which can be ascribed to Andreev reflections. The temperature and field dependence of the vdWH are in accordance with the BCS theory. Our results represent a probe-of-concept of the assembly of strongly correlated low dimensional materials in between 2D superconductors and open the doors to further studies involving 2D magnets or topological insulators as barriers. New exotic physical properties such as triplet superconductivity or Majorana fermions may arise from such combinations.

## 4. Methods

*Crystal growth*: High quality crystals of 2H-NbSe$_2$,[34] 2H-TaS$_2$[15] and 1T-TaS$_2$[21] are grown by Chemical Vapor Transport (CVT) using iodine as a transport agent, as already reported by some of us.

*vdWHs fabrication:* Bulk crystals are mechanically exfoliated and placed on top of 285 nm SiO$_2$/Si substrates using adhesive tape (80 μm thick adhesive plastic film from Ultron Systems). As a fast tool for the identification of thin-layers, the flakes are examined by optical microscopy (NIKON Eclipse LV-100 optical microscope under normal incidence). Atomic force microscopy images are taken with a Nano-Observer AFM from CSI Instruments. The heterostructures are built on top of pre-lithographed electrodes (5 nm Ti/50 nm Pd on 285 nm SiO$_2$/Si by NOVA Electronic Materials, LCC) by the deterministic assembly of the flakes using polycarbonate, as reported in reference [19], with the help of a micromanipulator. The whole process is performed inside an argon glovebox.

*Electrical measurement set-up:* Electrical measurements are performed in a Quantum Design PPMS-9 cryostat with a 4-probe geometry, where a DC current is passed by the outer leads and the DC/AC voltage drop is measured in the inner ones. DC voltage and AC differential resistance (27.7 Hz) are measured by conventional DC and AC lock-in techniques (the AC voltage is driven on top of the DC voltage and its value is the 0.1% of the maximum DC voltage) with a MFLI Lock-In Amplifier from Zurich Instruments, using an external resistance of 1 MΩ, i.e., much larger resistances than the sample. Field sweeps are performed at 200 Oe/s and temperature sweeps at 1K/min.

**Supporting Information**

Supporting Information is available from the Wiley Online Library or from the author.

**Acknowledgements**

We acknowledge the financial support from the European Union (ERC AdG Mol-2D 788222 and COST Action MOLSPIN CA15128), the Spanish MICINN (MAT2017-89993-R, RTI2018-098568-A-I and EQC2018-004888-P co-financed by FEDER and Excellence Unit "María de Maeztu", CEX2019-000919-M), and the Generalitat Valenciana (PO FEDER Program, ref. IDIFEDER/2018/061). R.C. acknowledges the support of a fellowship from "la Caixa" Foundation (ID 100010434). The fellowship code is LCF/BQ/PR19/11700008. C.B.-C. thanks the Generalitat Valenciana for a PhD fellowship. We thank Ángel López-Muñoz for his constant technical support and helpful discussions.

Received: ((will be filled in by the editorial staff))
Revised: ((will be filled in by the editorial staff))
Published online: ((will be filled in by the editorial staff))